\documentclass[12pt,a4paper]{article}
\usepackage{amsmath}
\usepackage{graphicx}
\usepackage{amssymb}
\usepackage{here}
\usepackage{tikz}
\usepackage{bm}
\usetikzlibrary{matrix}
\usepackage[normalem]{ulem}
\usepackage{overcite} 
\usepackage{dcolumn} 
\newcolumntype{.}{D{.}{.}{1}}

\begin{document}
\def\orbud{\scalebox{0.75}{\sout{$\;\upharpoonleft \,\downharpoonright\;$}}}
\def\orbu{\scalebox{0.75}{\sout{$\;\;\upharpoonleft \;\;$}}}
\def\orbd{\scalebox{0.75}{\sout{$\;\;\downharpoonleft \;\;$}}}
\def\orbe{\scalebox{0.75}{\sout{$\;\;\;\;$}}}

\newcommand{\rhor}{\ensuremath{\rho(\mathbf{r})}}
\newcommand{\mr}{\ensuremath{(\mathbf{r})}}
\newcommand{\dr}{\ensuremath{{\rm d}}}
\newcommand\Item[1][]{%
  \ifx\relax#1\relax  \item \else \item[#1] \fi
  \abovedisplayskip=0pt\abovedisplayshortskip=0pt~\vspace*{-\baselineskip}}

\begin{center}

\vspace*{1cm}

{\LARGE\bf
Self-Adaptive Tensor Network States\\[1.5ex] with Multi-Site Correlators
}

\vspace{1cm}

{\large Arseny Kovyrshin and Markus Reiher\footnote{Corresponding author: markus.reiher@phys.chem.ethz.ch}}
\\[2ex]
ETH Z\"urich, Laboratorium f\"ur Physikalische Chemie, 
Vladimir-Prelog-Weg 2, \\
8093 Z\"urich, Switzerland \\[2ex]

\begin{abstract}
We introduce the concept of self-adaptive tensor network states (SATNS) based on multi-site correlators. The SATNS ansatz 
gradually extends its variational space incorporating the most important next-order correlators into the ansatz for 
the wave function. The selection of these correlators is guided by entanglement-entropy measures from quantum information 
theory. By sequentially introducing variational parameters and adjusting them to the system under study, the SATNS ansatz achieves 
to keep their number significantly smaller than the total number of full-configuration interaction parameters. The 
SATNS ansatz is studied for manganocene in its lowest-energy sextet and doublet states, the latter of which is known to be
difficult to describe. It is shown that the SATNS parametrization solves the convergence issues found for previous
correlator-based tensor network states.
\end{abstract}

\vfill

\end{center}


\newpage
\section{Introduction}
The Density-Matrix Renormalization Group (DMRG)\cite{whit1992,scho2011} has become a reference approach in the electronic structure theory 
for problems with strong static electron correlation\cite{lege2008,chan2008a,chan2009a,reih2010,marti2011,chan2011,wout2014a,kura2014,yana2015,szal2015} 
such as open-shell transition metal complexes\cite{mart2008,bogu2012a,kura2013,shar2014,phun2016}. Its benefit is the polynomial scaling of computational 
costs due to a tensor-decomposition ansatz instead of the exponential scaling of traditional complete active-space approaches. Originally invented for 
one-dimensional spin chains\cite{whit1992} with nearest-neighbor interactions, the DMRG optimization algorithm for the full Coulomb problem inherited 
the one-dimensional order as a sequence of orbitals. 

Specifically, DMRG variationally optimizes a matrix product state (MPS) ansatz, a one-dimensional chain of tensors (tensor train)\cite{ostl1995,hackbusch_book}. 
The MPS ansatz allows for a compact description of entanglement in one-dimensional systems. Chemical systems are, however, governed by the full 
Coulomb interaction producing multidimensional entanglement. In comparison to one-dimensional systems, the convergence of DMRG is slower for 
chemical systems and a much larger bond-dimension must be taken into account. These problems can be partially alleviated either by employing 
optimized orbital ordering\cite{chan2002,lege2003,mori2005,lege2008,barc2011} on the sites of the DMRG lattice or by performing an orbital transformation\cite{krum2016}. 

In order to find an optimal wave function ansatz for the representation of multidimensional entanglement, a generalization of the MPS structure 
was necessary. As a consequence, a new family of wave function parameterizations emerged, the so-called tensor network 
states (TNS). Examples are the projected entangled pair states (PEPS)\cite{vers2004a}, tree tensor network states 
(TTNS)\cite{shi2006,tagl2009,corb2009,murg2010,barc2011,naka2013,murg2015}, the multiscale entanglement renormalization ansatz (MERA),\cite{even2014} and 
the wave function approximation based on the Tucker tensor decomposition\cite{mayh2017}.

The complete-graph TNS (CGTNS) ansatz\cite{mart2010b} adopts the correlator product state (CPS) ansatz\cite{gend2002,gend2003,mezz2009,chan2009}. The CGTNS 
ansatz allows all sites to interact with each other on equal footing. However, in its simplest form, it splits the interaction between sites into a product 
of pair correlators. The number of variational parameters in such a case depends only on the number of spin orbitals and scales as $\mathcal{O}(M^2)$, where 
$M$ is the number of spin orbitals. But as this ansatz considerably limits the variational degrees of freedom in a fixed fashion, it cannot adjust to a system under 
study to guarantee a homogeneous error. In our previous work\cite{kovy2016a} we increased the number of variational parameters employing 3-site correlators 
between sites (spin-orbitals) instead of 2-site correlators in the original CGTNS ansatz\cite{mart2010b}. Although the accuracy is then improved, 
the variational space is inflated, introducing even unnecessary degrees of freedom. In turn, this inflation leads to problems with the optimization of the ansatz\cite{kovy2016a}. 
Exploiting entanglement measures from quantum information theory\cite{lege2003,lege2004,riss2006}, we here propose to introduce extra variational freedom 
only for spin orbitals which are considered more entangled than others. We turn the CGTNS ansatz into a self-adaptive tensor network state (SATNS) 
ansatz which gradually evolves, guided by entanglement measures. The SATNS ansatz does not introduce all important higher-order correlators at a time but 
step-wise by starting from most important correlators. 

As our SATNS ansatz considers multi-site long-ranged correlators and the full Coulomb interaction, the deterministic optimization\cite{neus2011} 
of correlators is not feasible. Hence, for the optimization of correlators, a stochastic approach based on sampling in the discrete 
Slater determinant space is utilized\cite{mart2010b}. In this regard the developed method relates to Monte Carlo configuration interaction\cite{gree1995} 
(MCCI), which stochastically searches for important determinants in the full configuration interaction (FCI) space and to the very successful 
FCI Quantum Monte Carlo\cite{boot2009}.

\section{Theory}
In this section, we briefly review the original CGTNS ansatz and discuss its variants including higher-order correlators. In Subsection \ref{sec:entangl}, the 
concept of orbital entanglement entropy\cite{lege2003,lege2004} is introduced for this ansatz. Orbital entropies will play a key role in the extension of 
the self-adaptive tensor network described in Subsection \ref{sec:adaptive}. In the last Subsection \ref{sec:opt}, the stochastic optimization of the 
spin-adapted SATNS wave function is summarized, while a more detailed discussion of this algorithm can be found in Ref. \citen{kovy2016a}.

\subsection{Multi-Site Correlator Ansatz}\label{sec:Nsite}
The original CGTNS\cite{mart2010b} ansatz employs only 2-site correlators, which can be represented by second-order tensors for each pair of spin orbitals 
$i$ and $j$
\begin{equation}\label{eq:tensor2}
\mathbf{C}^{[ij]} \equiv \left[ 
\begin{array}{cc}
C_{00}^{[ij]} & C_{01}^{[ij]} \\
C_{10}^{[ij]} & C_{11}^{[ij]}
\end{array}
\right],
\end{equation}
where the row and column indices $n_i$ and $n_j$ of $C_{n_i n_j}^{[ij]}$ take only two values: 0 for an empty and 1 for an occupied spin orbital. Every 2-site 
correlator describes the entanglement between a pair of spin orbitals. Multiplying the elements of correlators corresponding to a certain occupation number 
vector (ONV), one approximates a state of $N$ electrons on $M$ spin orbitals by the 2-site CGTNS ansatz\cite{mart2010b} of the form
\begin{equation}\label{eq:CGTNS}
\left| \Psi^{2s} \right\rangle =\sum_{n_1 n_2 \ldots n_M} \prod_{i \le j} C_{n_i n_j}^{[ij]} 
\left| n_1 n_2 \ldots n_M \right\rangle.
\end{equation}
Eventually $M(M+1)/2$ correlators are used in the CGTNS ansatz, which makes the total number of variational parameters equal to $2M(M+1)$. For complete 
active space (CAS) -based methods the number of state parameters is defined by the number of ONVs (or configuration state functions for the 
spin-adapted case). As the active space is defined by the number of electrons, $N$, and a number of active spatial orbitals, $M_{\rm orb}=M/2$, we denote 
it as CAS($N$,$M_{\rm orb}$). If one considers only ONVs with the spin projection equal to zero and if no symmetry is exploited (not even particle conservation) 
then the total number of ONVs can be approximated as follows\cite{molcas8}
\begin{equation}\label{eq:cas}
N_{\rm ONV} \approx \frac{2}{\pi M_{\rm orb}}4^{M_{\rm orb}}=\frac{4}{\pi M}2^{M}.
\end{equation}
For systems with a number of active orbitals of less than eight, the CGTNS ansatz is able to reliably approximate this size of CAS wave 
function\cite{mart2010b}, but one will observe severe deviations for systems with larger active spaces. Such deficiencies may be cured by introducing 
higher-order correlators such as 3-site correlators\cite{kovy2016a} represented by a tensor of third order
\begin{equation}\label{eq:3-site}
\mathbf{C}^{[ijk]} \equiv \left[
\begin{tikzpicture}[baseline={([yshift=-.5ex]current bounding box.center)},vertex/.style={anchor=base,
    circle,fill=black!25,minimum size=18pt,inner sep=2pt}]
  \matrix (m) [matrix of math nodes, row sep=0.25em,
    column sep=0.25em]{
    & C_{001}^{[ijk]}& & C_{011}^{[ijk]} \\
    C_{000}^{[ijk]} & & C_{010}^{[ijk]} & \\
    & C_{101}^{[ijk]} & & C_{111}^{[ijk]} \\
    C_{100}^{[ijk]} & & C_{110}^{[ijk]} & \\};
  \path[]
    (m-1-2) edge (m-1-4) edge (m-2-1)
            edge [densely dotted] (m-3-2)
    (m-1-4) edge (m-3-4) edge (m-2-3)
    (m-2-1) edge [-,line width=6pt,draw=white] (m-2-3)
            edge (m-2-3) edge (m-4-1)
    (m-3-2) edge [densely dotted] (m-3-4)
            edge [densely dotted] (m-4-1)
    (m-4-1) edge (m-4-3)
    (m-3-4) edge (m-4-3)
    (m-2-3) edge [-,line width=6pt,draw=white] (m-4-3)
            edge (m-4-3);
\end{tikzpicture}\right].
\end{equation}
Each of these tensors describes the entanglement between three spin orbitals $i$, $j$, and $k$. Multiplying the elements of these correlators according to the 
occupations of the ONVs defines the 3-site CGTNS ansatz, 
\begin{equation}\label{eq:3s}
\left| \Psi^{3s} \right\rangle =\sum_{n_1 n_2 \ldots n_M} \prod_{i \le j \le k} C_{n_i n_j n_k}^{[ijk]} \left| n_1 n_2 \ldots n_M \right\rangle.
\end{equation}
Naturally, such an ansatz achieves a higher accuracy in the energy because of the larger number of variational parameters\cite{kovy2016a}, $M(M+1)(M+2)4/3$ 
resulting from the 8 elements of the $M(M+1)(M+2)/6$ $\mathbf{C}^{[ijk]}$ correlators. The tedious optimization of 3-site correlators can be enhanced by 
starting from pre-optimized 2-site correlators in a hybrid ansatz\cite{kovy2016a}, either by optimizing the 3-site correlators as scaling factors,
\begin{equation}\label{eq:hybrid1}
\left| \Psi^{3s[2s]} \right\rangle =\sum_{n_1 n_2 \ldots n_M} \underbrace{\prod_{i \le j} C_{n_i n_j}^{[ij]}}_{\rm frozen}
\underbrace{\prod_{k \le l \le m} C_{n_k n_l n_m}^{[klm]}}_{\rm active} \left| n_1 n_2 \ldots n_M \right\rangle,
\end{equation}
or preferably by summation of the products of the correlators of different order,
\begin{equation}\label{eq:hybrid2}
\left| \Psi^{3s+[2s]} \right\rangle =\sum_{n_1 n_2 \ldots n_M} \bigg[ \underbrace{\prod_{i \le j} C_{n_i n_j}^{[ij]}}_{\rm frozen} 
+ \underbrace{\prod_{k \le l \le m} C_{n_k n_l n_m}^{[klm]}}_{\rm active} \bigg] \left| n_1 n_2 \ldots n_M \right\rangle.
\end{equation}
Note that the number of variational parameters scales with the highest-order tensor in the expansion. All above-listed parameterization strategies can be naturally 
continued to higher-order correlators yielding, for instance, $\Psi^{4s}$, $\Psi^{5s}$, $\Psi^{6s}$. Defining the order of correlators by $L$ and recalling from 
combinatorics that the number of $L$-combinations with repetitions from a set of size $M$\cite{richard_stanley,benjamin_arthur} one obtains for the number of correlators
\begin{equation}
N_{\rm cor}= 
\left(
\begin{array}{c}
M+L-1\\
L\\
\end{array}
\right)=
\frac{(M+L-1)!}{L!(M-1)!}
,
\end{equation}
while the number of variational parameters is 
\begin{equation}
N_{\rm var}= 
\left(
\frac{(M+L-1)!}{L!(M-1)!}
\right)2^{L}.
\end{equation}
For both, $M$ and $L$, growing, the number of variational parameters exceeds the maximum number of variational parameters defined by traditional CAS-based 
methods, Eq.~(\ref{eq:cas}), unless one limits $L$ to some small value. Therefore, assuming that there are certain ranges for the number of active orbitals, 
where $L$ is limited to some value, $L<<M$, one may achieve polynomial scaling, $\mathcal{O}(M^L)$, for the number of variational parameters with system size. 
\begin{figure}[H]
\caption{Scaling of the variational parameters with an increasing number of active spatial orbitals $M_{\rm orb}$ in various CGTNS parameterizations 
(introduced in Sections \ref{sec:Nsite} and \ref{sec:adaptive}) and CAS-based methods. In the blue-shaded region, the CGTNS parameterizations introduce 
more variational parameters than the exact solution.
\label{fig:scaling}}
\begin{center}
\includegraphics[scale=0.6]{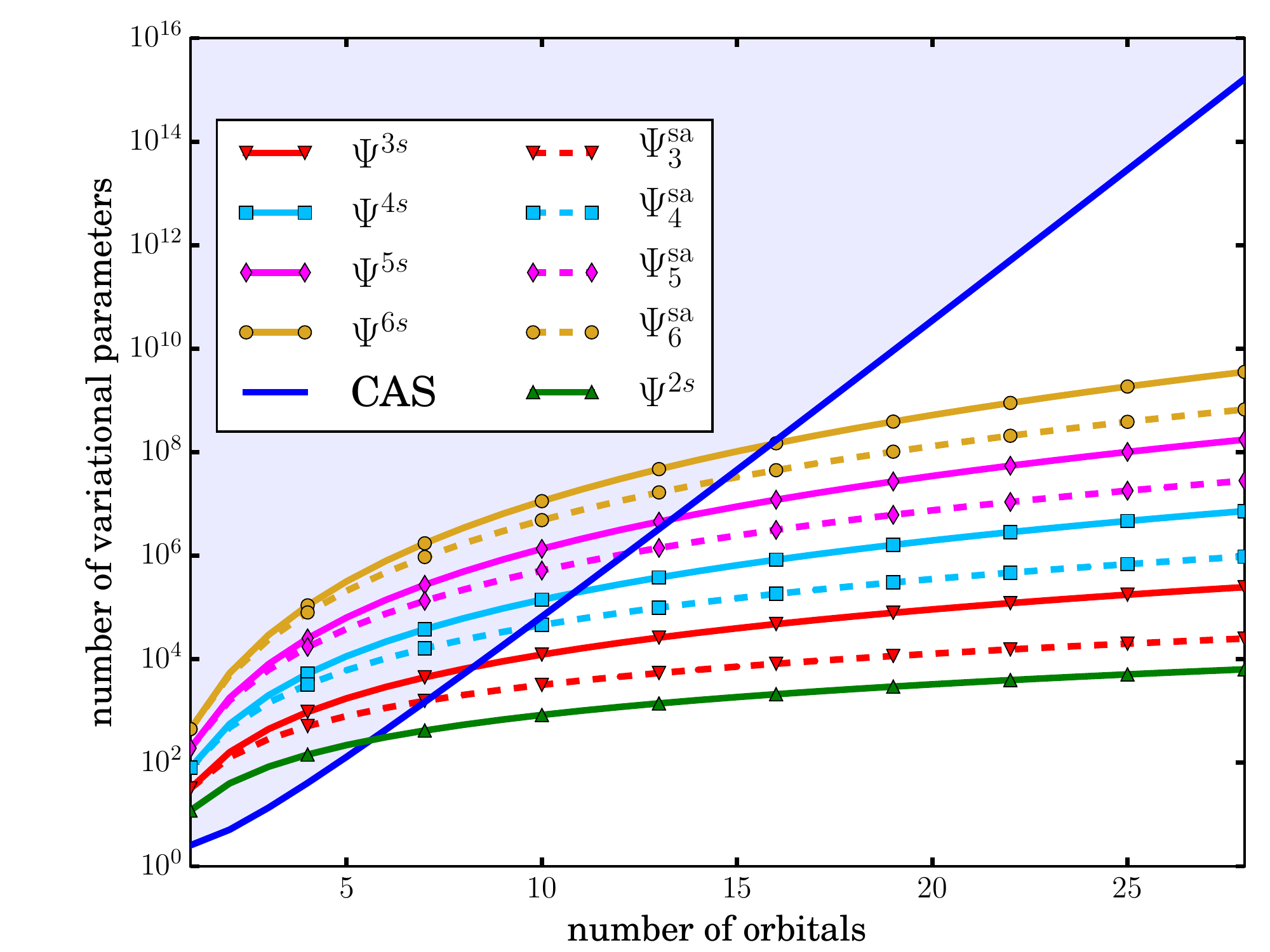}
\end{center}
\end{figure}
Figure~\ref{fig:scaling} shows the scaling of variational parameters with respect to the number of orbitals for various CGTNS parameterizations and CAS-based 
methods. However, whereas the 2-site correlators might, for example, already fail to yield an accurate energy, the number of variational parameters 
introduced in the 3-site correlator ansatz will, in general, be larger than needed, {\it i.e.}, larger than the one of a traditional CAS ansatz. As a solution 
to this problem, we suggested to limit the number of higher-order correlators and employ only the most important ones\cite{kovy2016a}.We now discuss how orbital 
entanglement may serve as a criterion for selecting the most important correlators.

\subsection{Entanglement Entropy}\label{sec:entangl}
For the definition of entanglement measures, we bipartition the CGTNS wave function by dividing the set of spin orbitals into an active part  
$\left| a \right \rangle = \left| a_{\alpha} a_{\beta} \right \rangle$ and an environment $\left| e \right \rangle$ ($e=\{n_1,n_2,\ldots,n_M\} 
\setminus \{a_{\alpha}, a_{\beta}\}$) 
\begin{equation}\label{eq:bipart}
\left| \Psi^{2s} \right\rangle = 
\sum_{a_{\alpha} \ldots e_{M-2}} 
\prod_{i \le j}^{\alpha,\beta} C_{a_i a_j}^{[ij]} 
\prod_{k}^{\alpha,\beta}
\prod_{l=1}^{M-2} C_{a_k e_l}^{[kl]} 
\prod_{m=1}^{M-2}
\prod_{m \le n}^{M-2} C_{e_m e_n}^{[mn]} 
\left| a_{\alpha} a_{\beta} e_1\ldots e_{M-2} \right\rangle. 
\end{equation}
The active part $\left| a \right\rangle$ may be one spatial orbital, 
\begin{equation}\label{eq:basis}
\{\left| a \right\rangle\}=\{\left| 0 \right\rangle\ , \left| {\rm up} \right\rangle\ , \left| {\rm down} \right\rangle\ , \left| 2 \right\rangle\},
\end{equation}
composed of $\alpha$- and $\beta$-spin orbitals located at the beginning of an ONV. 
For any other choice of the spatial orbital the resulting 
bipartition can be transformed into that of Eq.~(\ref{eq:bipart}) simply by permuting spin orbitals and taking care of sign changes. Introducing weights for 
each ONV $\left| a e \right\rangle = \left| a_{\alpha} a_{\beta} e_1\ldots e_{M-2} \right\rangle$ as
\begin{equation}
C_{ae}^{2s}= \left\langle a_{\alpha} a_{\beta} e_1 \ldots e_{M-2} \mid \Psi^{2s} \right\rangle = 
\prod_{i \le j}^{\alpha,\beta} C_{a_i a_j}^{[ij]} 
\prod_{k}^{\alpha,\beta}
\prod_{l=1}^{M-2} C_{a_k e_l}^{[kl]} 
\prod_{m=1}^{M-2}
\prod_{m \le n}^{M-2} C_{e_m e_n}^{[mn]}, 
\end{equation}
we can rewrite the bipartition in a more convenient way,
\begin{equation}
\left| \Psi^{2s} \right\rangle =
\sum_{ae} C_{ae}^{2s} \left| a \right\rangle  \left| e \right\rangle.
\end{equation}
Then, the density operator can be written as
\begin{equation}
\hat \rho =\sum_{ae} \sum_{a'e'} C_{ae}^{2s} C_{a'e'}^{2s*} \left| a \right \rangle \left| e \right\rangle 
\left\langle e' \right| \left\langle a' \right|,
\end{equation}
where we assume real expansion coefficients. After taking the partial trace over the environment, one obtains the reduced one-orbital density operator 
\begin{equation}
\hat \rho^{(1)} = {\rm Tr}_e \hat \rho = \sum_{e}  \left\langle e \right| \left( \sum_{ae} \sum_{a'e'} C_{ae}^{2s} C_{a'e'}^{2s*}\left| a \right\rangle \left| 
e \right\rangle \left\langle e' \right| \left\langle a' \right| \right) \left| e \right\rangle.
\end{equation}
Recalling the orthonormality of the basis $\{\left| e \right\rangle\}$ together with the fact that the number of particles and spin projection 
are conserved,\cite{riss2006,bogu2015} the expression for the one-orbital reduced density operator can be further simplified
\begin{equation}\label{eq:rho}
\hat \rho^{(1)} = \sum_{ae} \left| C_{ae}^{2s} \right|^2 \left| a \right\rangle \left\langle a \right|.
\end{equation}
Note that these definitions are easily extended to any multi-site CGTNS ansatz. Similar to the Shannon entropy in classical information 
theory, one can measure the information exchange between active part and environment, which characterizes the entanglement between 
them, with the von Neumann entropy
\begin{equation}\label{eq:vonNeuman}
s_{i}(1)=-{\rm Tr}( \hat \rho^{(1)} \ln \hat \rho^{(1)}).
\end{equation}
With a bipartition that puts only one spatial orbital into the active part this entropy is referred to as the single-orbital entropy\cite{lege2003,lege2004,riss2006}. 
Since the basis $\{\left| a \right\rangle\}$ is orthonormal, the (one-orbital) von Neumann entropy can be expressed as 
\begin{equation}\label{eq:orb_entropy}
s_{i}(1)=-\sum_{a}\omega_{a}\ln \omega_{a},
\end{equation}
where $a$ runs over the four possible occupations of the spatial orbital (0, up, down, 2) and $\omega_{a}=\sum_{e}\left| C_{ae}^{2s} \right|^2$.  
The single-orbital entropy measures the entanglement between orbital $a$ and the remaining set of the orbitals from the active space. In Eq.~(\ref{eq:bipart}) 
the correlators of the type $C_{a_{\alpha} n_i}^{[\alpha i]}$ and $C_{a_{\beta} n_i}^{[\beta i]}$ for all $i$ (of the environment) are responsible for the 
information exchange between the orbital $a$ and the rest of the system\cite{marti2011}. Hence, a high value of single-orbital entropy indicates that 
one needs to introduce additional variational degrees of freedom in order to better represent the entanglement of the orbital $a$  with the rest of the 
system. The natural choice is to introduce 3-site correlators of the type $C_{a_{\alpha} n_i n_j}^{[\alpha ij]}$ and $C_{a_{\beta} n_i n_j}^{[\beta ij]}$ 
for all possible $i$ and $j$, which satisfy the relation $i \le j$ (all belonging to the environment). In the following, we employ such sets of correlators for 
an extension of the 2-site CGTNS ansatz. Due to technical reasons these sets also contain the so-called self-interaction\cite{kovy2016a} correlators such as 
$C_{a_{\alpha} a_{\alpha} a_{\alpha}}^{\alpha \alpha \alpha}$ and $C_{a_{\alpha} a_{\alpha} n_j}^{\alpha \alpha j}$. 

\subsection{Self-Adaptive Tensor Network States}\label{sec:adaptive}
As a set of higher-order correlators might inflate the variational space up to unresolvable degree, it is important to limit them to certain sites selected based on 
the single-orbital entropies. In our previous work\cite{kovy2016a}, we showed that this efficiently decreases the variational space, simplifies the optimization problem, 
and yields accurate energies. Here, we suggest a self-adaptive strategy which introduces sets of higher-order correlators described in Section~\ref{sec:entangl} 
sequentially into the CGTNS ansatz. The algorithm at first optimizes the 2-site correlators and then after convergence evaluates single-orbital entropies. Afterwards, 
the set of 3-site correlators for the spin orbitals $\alpha$ $\beta$ corresponding to the orbital $a$ with the highest value of the single-orbital entropy is 
introduced into the ansatz, 
\begin{equation}\label{eq:ad}
\left| \Psi^{\rm sa} \right\rangle=
\sum_{n_1 \ldots n_M}
\left[
\underbrace{\prod_{i \le j}C_{n_i n_j}^{[ij]}}_{\rm frozen} +
\underbrace{\prod_{l \le m}\prod_k^{\alpha,\beta}P\left(C_{n_k n_l n_m}^{[klm]}\right)}_{\rm active}
\right]
\left| n_1 n_2 \ldots n_M \right\rangle,
\end{equation}
where the operator $P$ assures the order of the indices $k$, $l$, $m$ for correlators according to the following rule (since $k$ is not necessary at the beginning 
of the ONV)
\begin{equation}
P\left(C_{n_k n_l n_m}^{[klm]}\right) = 
\begin{cases}
C_{n_k n_l n_m}^{[klm]}, & \text{if } k \le l \le m\\
C_{n_l n_k n_m}^{[lkm]}, & \text{if } l \le k \le m\\
C_{n_l n_m n_k}^{[lmk]}, & \text{if } l \le m \le k 
\end{cases}\text{  .}
\end{equation}
Due to the fact that one of the indices ($k$) of 3-site correlators is forced to take only two values ($\alpha$ and $\beta$) the number of 3-site 
correlators scales similar to the number of 2-site correlators and is equal to
\begin{equation}
N_{\rm cor}= 
\left[
2\left(
\begin{array}{c}
M+1\\
2\\
\end{array}
\right)
-
M
\right],
\end{equation}
which yields 
\begin{equation}
N_{\rm var}= 
\left[
2\left(
\begin{array}{c}
M+1\\
2\\
\end{array}
\right)
-
M
\right]
8
\end{equation}
3-site variational parameters. Then, after freezing the optimized 3-site correlators, the algorithm automatically introduces correlators for the orbital with 
the second largest value of the single-orbital entropy, and so forth, generating new sets of correlators until no further decrease of the electronic
energy is observed. If, however, 3-site correlators introduced in this way always reduce the electronic energy to a non-negligible degree,
4-site correlator sets can be introduced. If no 3-site correlator set reduces the energy, a 4-site correlator set may be probed to ensure that
2-site correlators are sufficient and can be further optimized.

In the following, the set of correlators being optimized is referred to as the active set, while the associated orbital is called the active orbital. If the 
SATNS scheme uses correlators up to third order, it will be denoted as $\Psi^{\rm sa}_{3}$. One can also extend it with 4- and 5-site correlators and so 
forth which corresponds to $\Psi^{\rm sa}_{4}$, $\Psi^{\rm sa}_{5}$, and, ultimately, $\Psi^{\rm sa}_{L}$. For the general case, the number of variational parameters is
\begin{equation}
N_{\rm var}= 
\left[
2\left(
\begin{array}{c}
M+L-2\\
L-1\\
\end{array}
\right)
-
\left(
\begin{array}{c}
M+L-3\\
L-2\\
\end{array}
\right)
\right]
2^{L}.
\end{equation}
The SATNS introduces and optimizes the higher-order correlators systematically in such a way that the number of variational parameters within the active set of 
correlators has a similar scaling to that of the previous-order correlator ansatz, $\mathcal{O}(M^{L-1})$, instead of $\mathcal{O}(M^L)$ for all higher-order 
correlators introduced at once, see Figure~\ref{fig:scaling}.

\subsection{Optimization Algorithm}\label{sec:opt}
Within a given active space and one-particle basis set the expectation value of the electronic Hamiltonian $H$ over the SATNS ansatz can be
considered an upper bound to the 
complete active space configuration interaction (CAS-CI) reference energy,
\begin{equation}
E_{\rm CAS\text{-}CI}=\frac{\left\langle \Psi_{\rm CAS\text{-}CI} \right| H \left| \Psi_{\rm CAS\text{-}CI} \right\rangle}{\left\langle \Psi_{\rm CAS\text{-}CI} 
| \Psi_{\rm CAS\text{-}CI} \right\rangle} \le E_{\rm SATNS} = \frac{\left\langle \Psi_{L}^{\rm sa} \right| H \left| \Psi_{L}^{\rm sa} \right\rangle}
{\left\langle \Psi_{L}^{\rm sa} | \Psi_{L}^{\rm sa} \right\rangle}.
\end{equation}
For a spin-adapted CAS-CI wave function a linear expansion of spin-adapted configuration state functions (CSFs), 
$\left| \Phi^{\rm CSF}_{p} \right\rangle$, is employed
\begin{equation}\label{eq:sa}
\left| \Psi_{\rm CAS\text{-}CI} \right\rangle 
= \sum_{p} S_{p} \left| \Phi^{\rm CSF}_{p} \right\rangle
= \sum_{p} S_{p} \sum_{n} K_{pn} \left| n \right\rangle,
\end{equation}
where $K_{pn}$ are Clebsch--Gordan coefficients. Following our previous work\cite{kovy2016a}, we approximate
in the spin-adapted SATNS $S_{p}$ as 
follows
\begin{equation}\label{eq:st}
S_{p} \approx  S_{p}(\mathbf{\tilde{C}})= \sum_{n} K_{pn} C_{n}^{\rm sa},
\end{equation}
where $ C_{n}^{\rm sa}=\left(\prod_{i \le j} C^{[ij]}_{n_in_j} + \prod_{l \le m}\prod_k C_{n_k n_l n_m}^{[klm]}+ \ldots \right)$ for every $| n \rangle$ utilized in 
$\left| \Phi^{\rm CSF}_{p} \right\rangle$ and $\mathbf{\tilde{C}}$ defines a set of correlators $\{\mathbf{C}^{[11]}, \mathbf{C}^{[12]},$ $ \dots,$ 
$ \mathbf{C}^{[ij]},$ $ \dots,$ $ \mathbf{C}^{[NN]}\}$ constituting $C_{n}^{\rm sa}$. With the weights $S_{p}(\mathbf{\tilde{C}})$ from Eq.\ (\ref{eq:st}), 
we define the spin-adapted TNS ansatz as
\begin{equation}\label{eq:csf_like}
\mid \Psi^{\rm sa}_{L} \rangle = \sum_{p} S_{p}(\mathbf{\tilde{C}}) \mid \Phi^{\rm CSF}_{p} \rangle .
\end{equation}
The most efficient way to optimize correlators so far appeared to be the variational Monte Carlo scheme\cite{mart2010b,sand2007}, 
for which it is useful to rewrite the expression for $E_{\rm SATNS}$ as
\begin{equation}
E_{\rm SATNS}=\frac{\sum_{r}S_{r}(\mathbf{\tilde{C}})^{2} E_r(\mathbf{\tilde{C}})}
{\sum_{pq} S_p(\mathbf{\tilde{C}}) S_q(\mathbf{\tilde{C}}) \sum_{n} K_{pn} K_{qn}},
\end{equation}
where $S_{r}(\mathbf{\tilde{C}})^{2}$ is a probability distribution for energy estimators
\begin{equation}\label{eq:ener_estimate}
E_r(\mathbf{\tilde{C}})= \sum_{s} \frac{S_{s}(\mathbf{\tilde{C}})}{S_{r}(\mathbf{\tilde{C}})}\left\langle \Phi_{s}^{\rm CSF} 
\left| H \right| \Phi_{r}^{\rm CSF} \right\rangle.
\end{equation}
Having introduced an artificial thermal energy $T$ (measured in Hartree), the continuous variables $\mathbf{\tilde{C}}$ are sampled following a 
canonical ensemble with the configuration weights defined by $\exp{\left[-E_r(\mathbf{\tilde{C}})/T\right]}$. For higher efficiency, the parallel 
tempering scheme is applied\cite{mart2010b} during the optimization with swap-move probabilities between two neighboring temperatures defined as
\begin{equation}\label{eq:swap_prob}
p((T_i,E_i)\leftrightarrow(T_{i+1},E_{i+1}))={\rm min}\{1,\exp{(\Delta E/ \Delta T)}\},
\end{equation}
where $\Delta E = E_{i+1} - E_i$ and $\Delta T = T_{i+1}T_i/(T_i-T_{i+1})$. Each $T_l$ in the range $[T_1,T_P]$ is defined through the formula
\begin{equation}
T_{l}=T_{1}\left( \exp{\frac{\ln{T_{P}} - \ln{T_{1}}}{P-1}} \right)^{l-1}, \mbox{ with } l=1 \ldots P.
\end{equation}
Clearly an application of this optimization strategy is limited by the exponential scaling of the Hilbert space. 
Hence, it will only be feasible to work in a large CAS if not all CSFs are 
required so that those which hardly contribute can be omitted.

\begin{figure}[H]
\caption{Natural orbitals of manganocene in the lowest-energy sextet state that constitute the active spaces in the CAS(9,12)-SCF reference 
reproduced from the data published in Ref.~\citen{kovy2016a} ($C_{2v}$ symmetry).
\label{fig:sextets}}
\begin{center}
\includegraphics[scale=0.525]{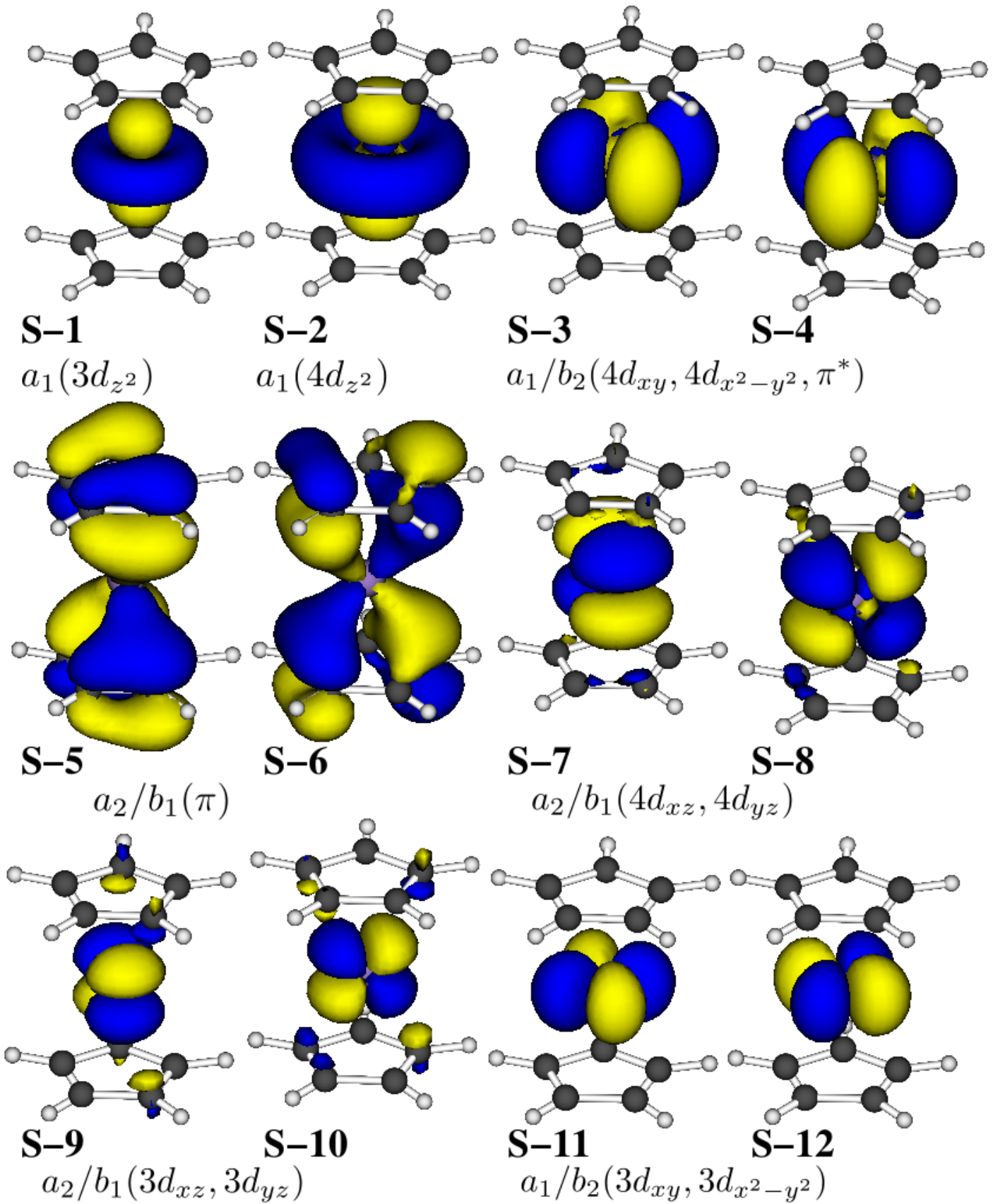}      
\end{center}
\end{figure}

\section{Computational Details}\label{sec:comp_det}
\begin{figure}[H]
\caption{Natural orbitals for manganocene in the lowest-lying doublet state that constitute the active spaces in the CAS(9,12)-SCF reference 
reproduced from the data published in Ref.~\citen{kovy2016a} ($C_{2v}$ symmetry).
\label{fig:doublets}}
\begin{center}
\includegraphics[scale=0.525]{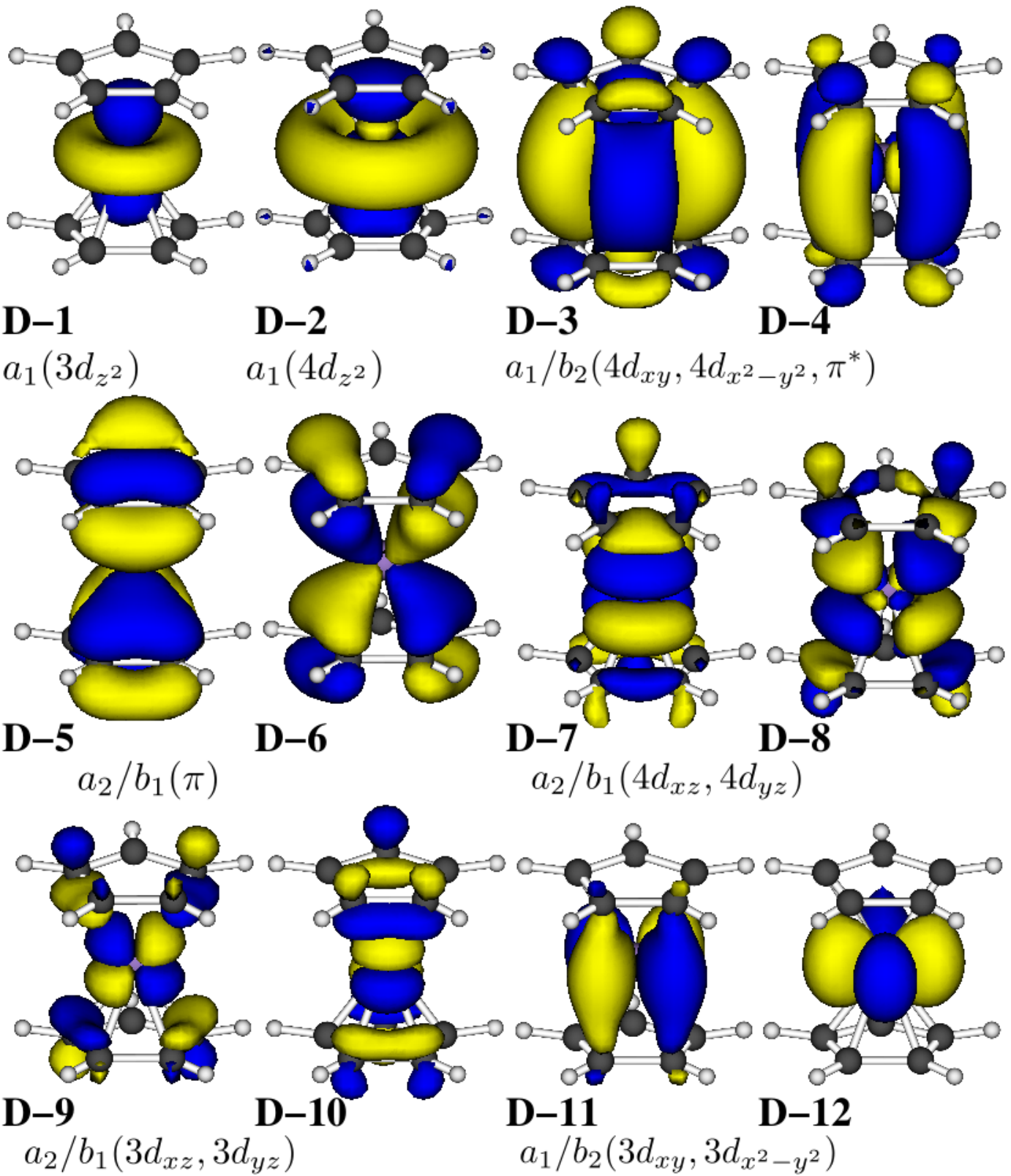}      
\end{center}
\end{figure}

To demonstrate the efficiency of the self-adaptive scheme compared to the multi-site CGTNS approach, we consider the manganocene complex, which 
served as an example in our previous work\cite{kovy2016a}. The structures of manganocene in doublet and sextet states as well as the doublet and 
sextet CAS(9,12)-SCF reference energies were taken from that previous work\cite{kovy2016a}. In order to introduce orbital labels for the discussion 
below, the converged natural orbitals from the CAS(9,12)-SCF calculation in $C_{2v}$ symmetry reported in Ref.~\citen{kovy2016a} for the sextet 
state ($^6A_1$) are shown in Figure~\ref{fig:sextets} and for the doublet ($^2A_1$) in Figure~\ref{fig:doublets}. For the second-quantized electronic 
Hamiltonian, the SATNS calculations employ one- and two-electron molecular orbital integrals generated from these natural orbitals. We refer to 
Ref.~\citen{kovy2016a} for details on how these orbitals and energies were obtained.

In our CGTNS program\cite{kovy2016a}, we implemented the single-orbital entropies  and the self-adaptive ansatz, $\Psi^{\rm sa}_{n}$, discussed in 
Sections \ref{sec:entangl} and \ref{sec:adaptive}, respectively. The reference values of single-orbital 
entropies were taken from MPS-DMRG calculations carried out with {\sc QCMaquis}\cite{kell2016}, which employed the natural orbitals from the reference 
CAS-SCF wave functions. The number of renormalized block states in these DMRG calculations
was set to 1000, while the maximum number of sweeps was limited to 20. 

The parallel tempering Monte Carlo scheme described in Section~\ref{sec:opt} was employed for optimization of SATNS ansatz. The extension of the SATNS 
ansatz with a new set of correlators was made every 20 Monte Carlo steps. This, in general, appeared to be enough for reaching convergence within the 
active set of correlators, although a more rigorous control for changing to the new correlator set can be based on the convergence of the energy 
and the single-orbital entropies. Note also that correlators at the beginning of the optimization, when SATNS ansatz lacks variational freedom, might 
fail to adequately describe the true wave function. Hence, the thorough relaxation of correlators  may converge the ansatz to a wrong solution, which would 
create additional difficulties in the optimization of higher-order correlators. For the same reason, our SATNS ansatz performs quick pre-optimization 
(20 Monte Carlo steps) of the 2-site correlators instead of employing the fully relaxed 2-site correlators from the $\Psi^{2s}$ ansatz (as in the 
case of $\Psi^{3s+[2s]}$ ansatz). However, once all necessary higher-order correlators are introduced one can perform a second optimization round. 
{\it I.e.}, the algorithm continues with a refinement of the SATNS ansatz starting over from 2-site correlators and switching to subsequent sets of correlators 
in the same order as it was done in the first optimization round. As will be shown in Section~\ref{sec:sextet} this might lower the energy further.
If the improvement is significant, the algorithm can continue performing optimization rounds until convergence is achieved.

\section{Results}
In this section, we calculate the adiabatic doublet--sextet electronic energy splitting of the manganocene spin-crossover 
complex with our SATNS ansatz. 
As can be understood from our previous results\cite{kovy2016a}, the electronic states of manganocene represent a challenging test for electronic
structure methods in general and for CGTNS-type schemes in particular. 
The reference CAS(9,12)-SCF data, 
the data for 2- and 3-site CGTNS parameterizations ($\Psi^{2s}$ and $\Psi^{3s}$), as well as for the hybrid ansatz, $\Psi^{3s[2s]}$, were taken from the 
Ref.~\citen{kovy2016a}.

\subsection{Manganocene sextet}\label{sec:sextet}
The CAS(9,12)-SCF wave function for manganocene in its lowest-energy sextet state features a rather small variational space of 13108 ONVs that can be well approximated 
with the 2-site correlator ansatz\cite{kovy2016a}. Such an approximation eliminates 91 \% of the CAS-CI variational space introducing an error of around 15 mHartree, 
see Table~\ref{tab:sextet}. Unfortunately, if one employs solely 3-site correlators through the $\Psi^{3s}$ ansatz or in a hybrid scheme together with the 
2-site correlators as in the $\Psi^{3s[2s]}$ ansatz, this does not improve the energy significantly and yields in the best case only -1542.197777 Hartree. Together 
with the fact that the variational space is increased by $59$ \% in this case, the ansatz is rendered useless. 

\begin{table}[H]
\caption{Electronic energies for the sextet state of manganocene calculated with various CGTNS-type parameterizations, SATNS, and CAS(9,12)-SCF.
}\label{tab:sextet}
\begin{center}
%
\begin{tabular}{lrl}
\hline\hline
parameterization & parameters & energy/Hartree \\
\hline
\multicolumn{3}{c}{first optimization round}   \\
\hline
 CAS-SCF             & 13108  & $-1542.209620$\textsuperscript{a} \\    
 $\Psi^{\rm sa}_{3}$ & 4608   & $-1542.209585$                    \\    
 $\Psi^{3s}$         & 20800  & $-1542.197777$\textsuperscript{a} \\    
 $\Psi^{3s[2s]}$     & 20800  & $-1542.195283$\textsuperscript{a} \\    
 $\Psi^{2s}$         & 1200   & $-1542.194072$\textsuperscript{a} \\    
\hline
\multicolumn{3}{c}{second optimization round}   \\
\hline
 $\Psi^{\rm sa}_{3}$ & 4608  &  $-1542.209603$ \\    

\hline\hline
\multicolumn{3}{l}{\textsuperscript{a}\footnotesize{these electronic energies were taken from Ref.~\citen{kovy2016a}.}}
\end{tabular}
\end{center}
\end{table}

\begin{figure}[H]
\caption{Convergence behavior of the $\Psi^{\rm sa}_{3}$, $\Psi^{3s}$, and $\Psi^{3s[2s]}$ parameterizations for manganocene in the lowest-lying sextet state. The data 
for $\Psi^{2s}$, $\Psi^{3s}$, $\Psi^{3s[2s]}$, and reference CAS(9,12)-SCF were taken from Ref.~\citen{kovy2016a}. Vertical gray dashed lines show the Monte Carlo 
steps in which new correlators were introduced. The vertical red dashed line indicates the Monte Carlo step in which the algorithm switched from 2-site to 3-site correlators.
\label{fig:conv_sext}}
\begin{center}
\includegraphics[scale=0.6]{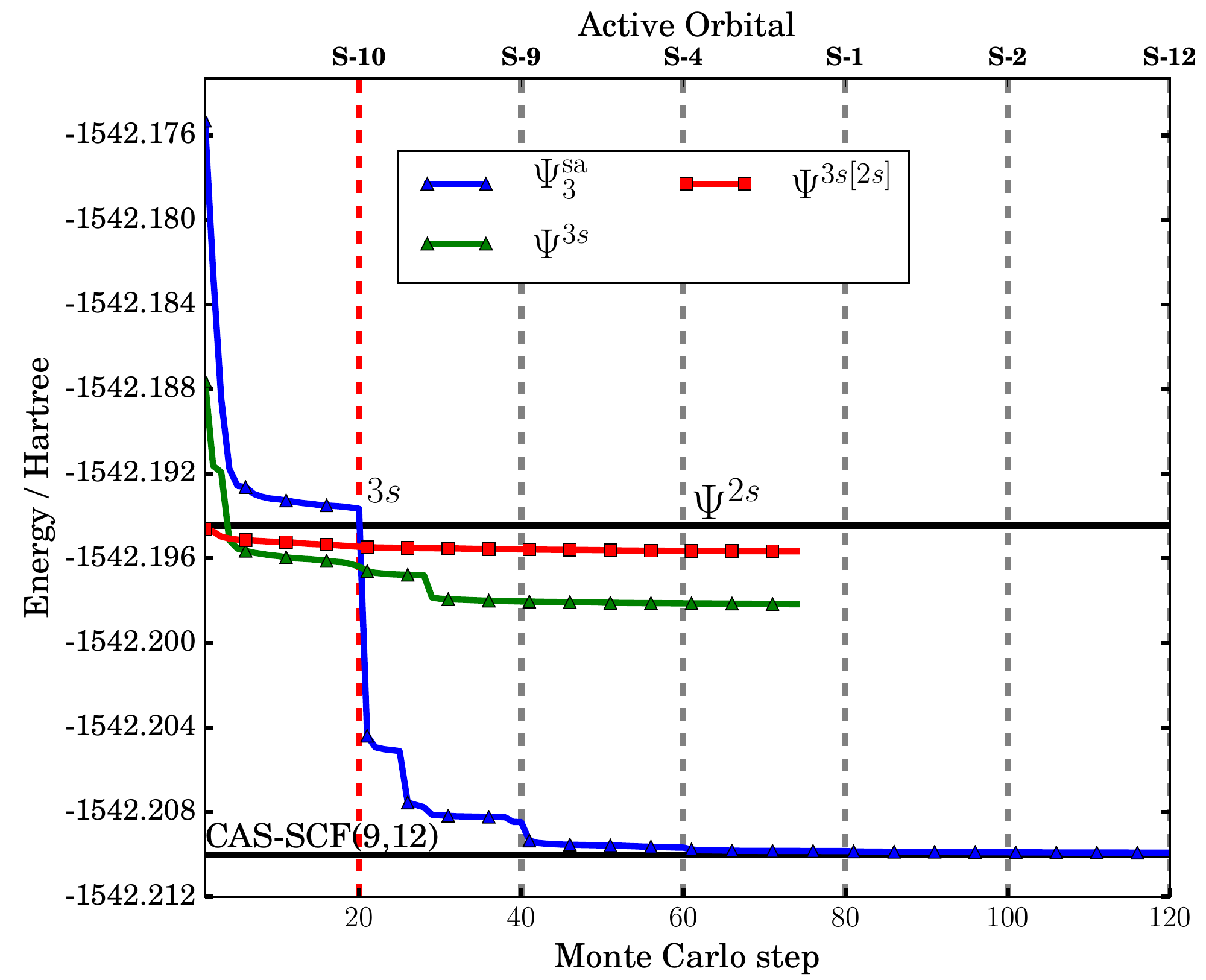}
\end{center}
\end{figure}

\begin{figure}[H]
\caption{Convergence behavior of single-orbital entropies (manganocene in the lowest-energy sextet state) for the $\Psi^{\rm sa}$ ansatz. 
The reference values of the single-orbital entropies are taken from a MPS-DMRG calculation and shown as horizontal black dashed lines. 
We identify five sets, each composed of near degenerate single-orbital entropies, namely $\{${\bf S-1},{\bf S-2}$\}$, $\{${\bf S-11},{\bf S-3},{\bf S-12},
{\bf S-4}$\}$, $\{${\bf S-9},{\bf S-10}$\}$, $\{${\bf S-7},{\bf S-8}$\}$, and $\{${\bf S-5},{\bf S-6}$\}$. Vertical gray dashed lines indicate the Monte Carlo 
steps in which new correlators were introduced. The vertical red dashed line shows the Monte Carlo step in which the order changed from 2-site to 3-site).
\label{fig:sextet_qi}}
\begin{center}
\includegraphics[scale=0.6]{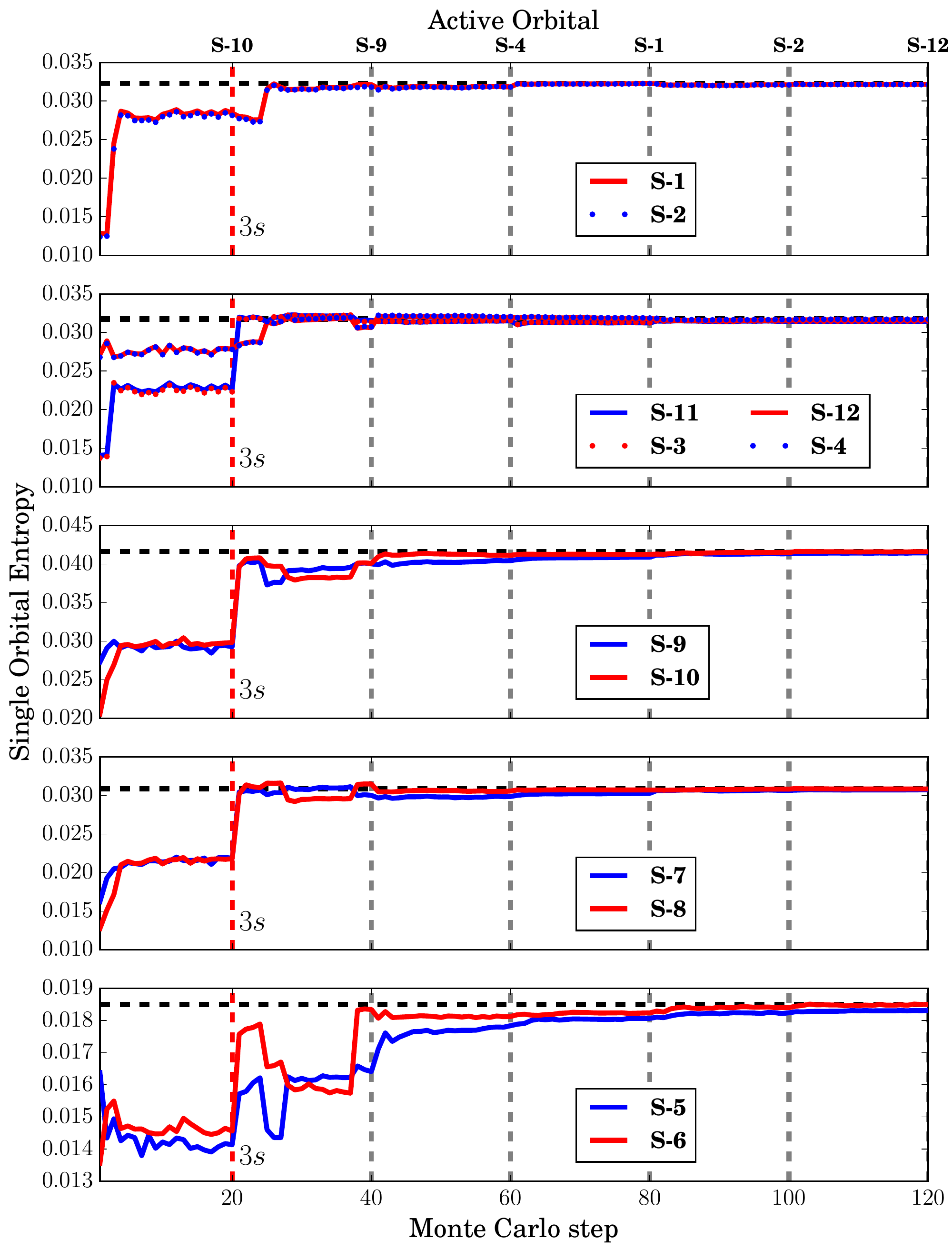}
\end{center}
\end{figure}

This problem is solved by the SATNS ansatz. As expected, after 20 Monte Carlo steps in the optimization of 2-site correlators, the energy of the SATNS ansatz (based solely on 
2-site correlators) does not reach the minimum of a $\Psi^{2s}$ ansatz, which is slightly lower (see Figure~\ref{fig:conv_sext}). In order to study the convergence 
of SATNS, we monitor the entanglement of the orbitals in Figure~\ref{fig:sextet_qi}, where the reference single-orbital entropies 
were taken from the MPS-DMRG calculation. The reference single-orbital entropies are degenerate and have only five different values 0.0323, 0.0317, 0.0416, 0.0309, 
0.0185 for sets $\{${\bf S-1}, {\bf S-2}$\}$, $\{${\bf S-11}, {\bf S-3}, {\bf S-12}, {\bf S-4}$\}$, $\{${\bf S-9}, {\bf S-10}$\}$, $\{${\bf S-7}, {\bf S-8}$\}$, 
$\{${\bf S-5},{\bf S-6}$\}$, respectively. One can see that the 2-site correlators poorly describe the entanglement between orbitals. In the region up to 
Monte Carlo step 20, the values of single-orbital entropies for all orbitals are far from the reference values and oscillate during the optimization. As 
the orbital {\bf S-10} has the largest single-orbital entropy the SATNS scheme introduces corresponding 3-site correlators according to Eq.~(\ref{eq:ad}) 
and one observes a large energy drop, see Figure~\ref{fig:conv_sext}. Note that only the first set of 3-site correlators provides such a pronounced
decrease in energy. Also, this set considerably improves the values of single-orbital entropies for almost all orbitals except for {\bf S-5} 
and {\bf S-6}. The entanglement for these orbitals approaches the reference only after the extension of the ansatz with the correlators for the orbitals {\bf S-9} 
and {\bf S-4}. For these correlator sets, one can again observe significant decreases in energy which brings the SATNS electronic energy
very close to the reference value. Introducing correlators 
gradually for the most entangled orbitals, the SATNS manages to bypass the optimization problems observed in Ref.\
\citen{kovy2016a} for the $\Psi^{3s[2s]}$ and $\Psi^{3s}$ 
parameterizations. In contrast to $\Psi^{3s}$, it reduces the variational space by 65 \%, optimizing only 4608 variational parameters at a time. The CPU time 
required for one Monte Carlo step in the optimization of 3-site correlators in $\Psi^{\rm sa}_{3}$ ansatz is only 1.46 times longer than 
the one for the $\Psi^{2s}$ ansatz, 
while in the case of $\Psi^{3s}$ and $\Psi^{3s[2s]}$ it is 4.77 times longer. If the 3-site correlators are introduced for all orbitals, the error in energy will 
be less than 0.1 mHartree, see Table~\ref{tab:sextet}. In this case, the single-orbital entropies closely approach the reference values, see Figure~\ref{fig:sextet_qi}. 
An extension of the ansatz with 4-site correlators only leads to a negligible decrease of the electronic energy, hence showing that the 2- and 3-site correlators alone are 
able to adequately approximate the wave function for manganocene in the sextet state. 

The same holds also true for the entanglement between the orbitals. Such a fast convergence of the SATNS wave function can, however, be attributed to the relatively simple 
entanglement pattern found for this electronic state. Note the rather small single-orbital entropies 
($0 < s_{i}(1) < 0.1$)\cite{bogu2012b}.
Although the 3-site-correlator SATNS ansatz $\Psi^{\rm sa}_{3}$ provides a significantly more accurate energy compared to the $\Psi^{3s}$ ansatz, one still 
can improve on it by letting all correlator sets relax further in a second optimization round. This yields an electronic energy of $-1542.209603$ Hartree, see 
Table~\ref{tab:sextet}.

\subsection{Manganocene doublet}
In contrast to the sextet state, the doublet state has a larger variational space of 
98000 ONVs with a more complicated entanglement pattern. The single-orbital entropies for the orbitals {\bf D-1}, {\bf D-12}, {\bf D-2}, 
{\bf D-3}, $\{${\bf D-5}, {\bf D-6}$\}$, {\bf D-4}, $\{${\bf D-10}, {\bf D-9}$\}$, {\bf D-11}, and $\{${\bf D-7}, {\bf D-8}$\}$ are 0.1976, 0.1311, 0.1676, 
$\{$ 0.2125, 0.2121$\}$, 0.2236, $\{$0.2320, 0.2316$\}$, 0.2489, $\{$0.0097, 0.0097$\}$, respectively (the orbitals in curly brackets have near-degenerate values of 
single-orbital entropies). 
All the orbitals except {\bf D-7} and {\bf D-8} have moderately large single-orbital entropies ($0.1 < s_{i}(1) < 0.5$)\cite{bogu2012b}. 

\begin{figure}[H]
\caption{Convergence behavior of the $\Psi^{\rm sa}_{4}$, $\Psi^{3s}$, and $\Psi^{3s[2s]}$ parameterizations for manganocene in the doublet state. The 
data for $\Psi^{2s}$, $\Psi^{3s}$, $\Psi^{3s[2s]}$, and reference CAS(9,12)-SCF were taken from Ref.~\citen{kovy2016a}. Vertical gray dashed 
lines show the Monte Carlo steps where new correlators are introduced. Vertical red dashed lines show the Monte Carlo steps where the order of new 
correlators changes ({\it e.g.}, from 2-site to 3-site).
\label{fig:conv_doub}}
\begin{center}
\includegraphics[scale=0.6]{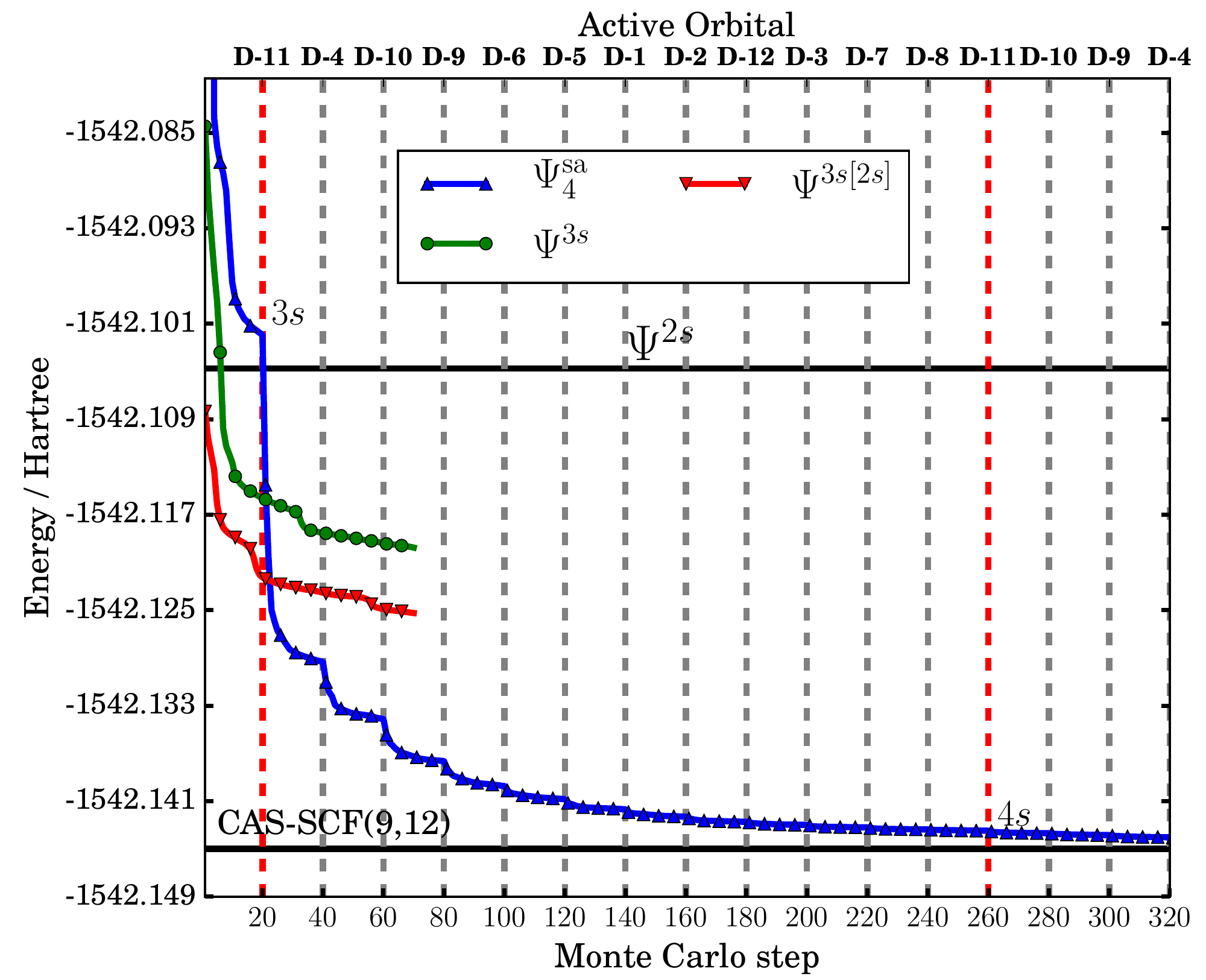}
\end{center}
\end{figure}

\begin{figure}[H]
\caption{Convergence behavior of single-orbital entropies (manganocene in the doublet state) for the self-adaptive $\Psi^{\rm sa}$ ansatz. 
The reference values of the single-orbital entropies are taken from a MPS-DMRG calculation are shown as horizontal black dashed lines. For the better readability 
the data were split into 5 sets: $\{${\bf D-1},{\bf D-12},{\bf D-2}$\}$, $\{${\bf D-3}$\}$, $\{${\bf D-5},{\bf D-6},{\bf D-4}$\}$, $\{${\bf D-10}, 
{\bf D-9},{\bf D-11}$\}$, and $\{${\bf D-7},{\bf D-8}$\}$. Some of the single-orbital entropies form near degenerate sets,namely $\{${\bf D-5},
{\bf D-6}$\}$, $\{${\bf D-10},{\bf D-9}$\}$, and $\{${\bf D-7},{\bf D-8}$\}$. Vertical gray dashed lines indicate the Monte Carlo steps in which new correlators 
were introduced. Vertical red dashed lines point to those Monte Carlo steps in which the order of new correlators changes ({\it e.g.}, from 2-site to 3-site).
\label{fig:doublet_qi}}
\begin{center}
\includegraphics[scale=0.6]{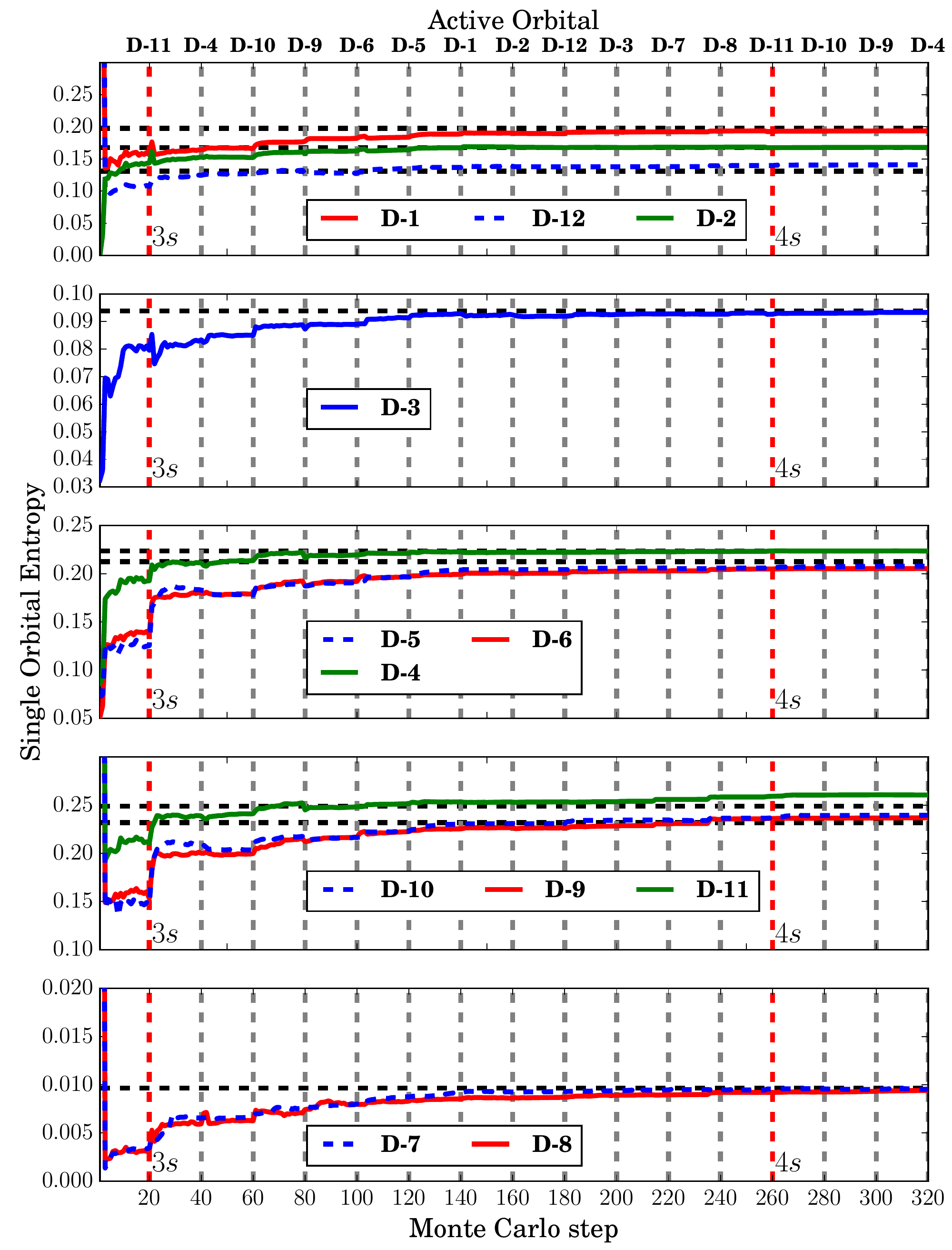}
\end{center}
\end{figure}

If the CGTNS parameterizations from Section~\ref{sec:Nsite} are considered, the most accurate value of the energy is obtained with the $\Psi^{3s[2s]}$ 
ansatz\cite{kovy2016a}. Still, the error of this parameterization is half of the energy difference between the reference value and the energy of the $\Psi^{2s}$ 
ansatz, while one Monte Carlo step is 4.81 times slower than one for the $\Psi^{2s}$ ansatz. The SATNS reaches the accuracy of the $\Psi^{3s[2s]}$ ansatz after an extension 
with the first set of 3-site correlators for the orbital {\bf D-11}, see Figure~\ref{fig:conv_doub}. The single-orbital entropies for all orbitals oscillate 
during the optimization and poorly approximate the reference wave function until the 3-site correlators are introduced, see Figure~\ref{fig:doublet_qi}. For 
the orbitals with a high value for the single-orbital entropy, namely {\bf D-10}, {\bf D-9}, {\bf D-11}, {\bf D-4}, {\bf D-5}, and {\bf D-6}, the entanglement 
is significantly underestimated. Also the entanglement pattern for these orbitals is considerably improved with the introduction of the first 3-site correlator 
set. The other orbitals, however, do not experience such an improvement. Their single-orbital entropies slowly converge to the reference 
values with an introduction of 3-site correlators for the rest of the most entangled orbitals, namely {\bf D-4}, {\bf D-10}, {\bf D-9}, {\bf D-6}, and {\bf D-5}. 
From the energy convergence shown in Figure~\ref{fig:conv_doub}, one can clearly see that these orbitals are also responsible for the major decrease 
of the electronic energy. 
With the introduction of 3-site correlators for the orbitals {\bf D-2}, {\bf D-12}, {\bf D-3}, {\bf D-7}, {\bf D-8}, the energy slowly decreases, but does not 
reach the same accuracy as for the case of the sextet yielding only $-1542.143410$ Hartree, see Table~\ref{tab:doublet2} and Figure~\ref{fig:conv_doub}. 

\begin{table}[H]
\caption{Electronic energies for the doublet state of manganocene calculated with various CGTNS-type parameterizations, SATNS, and CAS(9,12)-SCF.
}\label{tab:doublet2}
\begin{center}
\begin{tabular}{lrl}
\hline\hline
parameterization & parameters & energy/Hartree \\   
\hline
\multicolumn{3}{c}{first optimization round}   \\
\hline
 CAS-SCF               & 98060 &  $-1542.144937$\textsuperscript{a}\\
 $\Psi^{\rm sa}_{4}$   & 78400 &  $-1542.144416$                    \\   
 $\Psi^{\rm sa}_{3}$   & 4608  &  $-1542.143410$                    \\   
 $\Psi^{3s[2s]}$       & 20800 &  $-1542.125171$\textsuperscript{a} \\   
 $\Psi^{3s}$           & 20800 &  $-1542.119695$\textsuperscript{a} \\   
 $\Psi^{2s}$           & 1200  &  $-1542.104681$\textsuperscript{a} \\   
\hline\hline
\multicolumn{3}{l}{\textsuperscript{a}\footnotesize{these electronic energies were taken from Ref.~\citen{kovy2016a}.}}
\end{tabular}
\end{center}
\end{table}

The reduction in variational space achieved by the $\Psi^{\rm sa}_{3}$ ansatz is with 95\% high and one Monte Carlo step in the 
optimization of the set of 3-site correlators is almost as fast as for the $\Psi^{2s}$ ansatz, see Table~\ref{tab:doublet2}. 
Extending the self-adaptive ansatz 
toward 4-site correlators one observes a nonnegligible drop in energy, which shows the importance of 4-site correlators for the wave function in contrast to 
the sextet state. One Monte Carlo step in the optimization of the 4-site correlators in the $\Psi^{\rm sa}_{4}$ ansatz is 3.87 times slower in comparison to 
the $\Psi^{2s}$ ansatz. After the introduction of the 4-site correlators for all orbitals the energy for the doublet state becomes $-1542.144416$ Hartree, for 
which, however, 80 \% of the CAS-CI variational space is required. 

A problem occurs for the entanglement description of orbital {\bf D-11}. The corresponding single-orbital 
entropy, after reaching the reference value at Monte Carlo step 100, continues to grow and at Monte Carlo step 320 it is slightly overestimated. A 
possible reason might be an insufficient entanglement description in the SATNS ansatz at the beginning of the optimization: at Monte Carlo step 
40, the 3-site correlators were introduced for the {\bf D-4} orbital, as it had the second largest single-orbital entropy in that step. 
But the better choice should have been made in 
favor of the {\bf D-10} orbital since it has the second largest single-orbital entropy according to the reference wave function and the SATNS at later stages of 
an optimization.
If such issues are not properly resolved in not fully optimized SATNS wave functions, parallel or simultaneous 
optimization strategies may be considered.

\subsection{Sextet--doublet energy splitting}
To demonstrate the accuracy of our self-adaptive tensor network approach for the more relevant relative energies, 
we summarize the results of CGTNS parameterizations from our previous work\cite{kovy2016a}.
The worst approximation for the energy splitting is obtained with the $\Psi^{2s}$ ansatz, which yields -56.09 kcal/mol, see Table~\ref{tab:difference}. 
3-site correlators reduce the error relative to the CAS(9,12)-SCF reference from 16.5 to 8.41 kcal/mol. If 2-site correlators are combined 
with 3-site correlators in the hybrid $\Psi^{3s[2s]}$ scheme, the error is decreased further to 3.41 kcal/mol. The self-adaptive ansatz dramatically improves 
on those results: already a SATNS parameterization with correlators up to only third order reduces the error to 0.94 kcal/mol. 
Employing $\Psi^{\rm sa}_{3}$ for the sextet and $\Psi^{\rm sa}_{4}$ for the doublet yields an error of 0.31 kcal/mol.

\begin{table}[H]
\caption{The doublet--sextet energy differences in kcal/mol for manganocene calculated with various CGTNS parameterizations and CAS(9,12)-SCF.}
\label{tab:difference}
\begin{center}
\begin{tabular}{rrl}
\hline\hline
\multicolumn{2}{c}{parameterization}                            & \multicolumn{1}{c}{$E[^6A_1]-E[^2A_1]$} \\
 $^6A_1$ & $^2A_1$ & kcal/mol \\
\hline
\\
 \multicolumn{2}{c}{CAS-SCF}               &  $-40.59$\textsuperscript{a} \\
 $\Psi^{\rm sa}_{3}$ & $\Psi^{\rm sa}_{4}$ &  $-40.90$ \\
 \multicolumn{2}{c}{$\Psi^{\rm sa}_{3}$}   &  $-41.53$ \\
 \multicolumn{2}{c}{$\Psi^{3s[2s]}$}       &  $-44.00$\textsuperscript{a} \\
 \multicolumn{2}{c}{$\Psi^{3s}$}           &  $-49.00$\textsuperscript{a} \\
 \multicolumn{2}{c}{$\Psi^{2s}$}           &  $-56.09$\textsuperscript{a} \\
\hline\hline
\multicolumn{3}{l}{\textsuperscript{a}\footnotesize{taken from Ref.~\citen{kovy2016a}.}}
\end{tabular}
\end{center}
\end{table}

\section{Conclusions}
As the order of correlators is increased in the $L$-site CGTNS scheme, the fast growth of the variational space is a major problem. 
Moreover, the higher-order correlators make the optimization procedure much more involved. 
Here, we developed a self-adaptive strategy which overcomes these problems. Our SATNS strategy is based 
on the evaluation of single-orbital entropies, which has been introduced and implemented in this work for CGTNS-type parameterizations.
The SATNS ansatz starts with the optimization of 2-site correlators. Then, based on single-orbital entropies it gradually extends an ansatz with 
higher-order correlators for the most entangled orbitals. Thereby, a selective optimization rather than an optimization of all higher-order correlators
is achieved. Our manganocene case study showed that the SATNS 
is able to systematically improve the accuracy employing only a small part of the variational space arising from the complete set of correlators of 
a certain order. The $\Psi^{\rm sa}_{3}$ scheme achieves an accuracy with an error of about one kcal/mol, while about 65 \% of the variational space 
was reduced in the case of the sextet and 95 \% for the doublet. 
Various aspects of the SATNS parametrization should be studied in future work, of which excited electronic states and gradients are two examples.

\section{Acknowledgment}
This work was supported by ETH Zurich (ETH Fellowship FEL-27 14-1) and by the Schweizerischer Nationalfonds (Project No.\ SNF $200020\_169120$).


\end{document}